\documentclass[aps,pra,twocolumn,showpacs]{revtex4}

\usepackage[dvips]{graphicx}
\usepackage[normalem]{ulem}

\input{epsf}

\begin{document}

\title{Equation of state of an interacting Bose gas at finite temperature: \\
 a Path Integral Monte Carlo study}

\author{S. Pilati$^{(1)}$, K. Sakkos$^{(2)}$, J. Boronat$^{(2)}$, J. Casulleras$^{(2)}$, and S. Giorgini$^{(1)}$}
\affiliation{$^{(1)}$Dipartimento di Fisica, Universit\`a di Trento and CRS-BEC INFM, I-38050 Povo, Italy\\
$^{(2)}$Departament de F\'{\i}sica i Enginyeria Nuclear, Campus Nord B4-B5, Universitat Polit\`ecnica 
de Catalunya, E-08034 Barcelona, Spain}

\date{\today}

\begin{abstract}
By using exact Path Integral Monte Carlo methods we calculate the equation
of state of an interacting Bose gas as  a function of temperature both
below and above the superfluid transition. The universal character of the
equation  of state for dilute systems and low temperatures is investigated 
by modeling the interatomic interactions using different repulsive potentials 
corresponding to the same $s$-wave scattering length. The results obtained for 
the energy and the pressure are compared to the virial expansion for temperatures 
larger than the critical temperature. At very low temperatures we find agreement 
with the ground-state energy calculated using the diffusion Monte Carlo method.        
\end{abstract}


\maketitle

\section{Introduction}
\label{Introduction}

In the last decade, after the first realization of Bose-Einstein
condensation (BEC) in dilute systems of alkali  atoms~\cite{BEC}, the
experimental and theoretical investigation of quantum degenerate gases has
become one  of the most active and fast developing fields in atomic,
molecular and condensed matter physics~\cite{book}.  The effect of
interatomic interactions on the properties of ultracold Bose gases has been
the subject of a deep  and extensive research activity. As for the
theoretical side, the problem has been addressed from many   different
perspectives, both at zero and finite temperature, focusing on dynamical or
equilibrium properties,  in various dimensionalities and geometrical
configurations both homogeneous and inhomogeneous. Many different methods 
have also been used, from simple mean-field approaches to more advanced and
essentially exact quantum Monte Carlo  techniques~\cite{book,review}. In
particular, the Path Integral Monte Carlo (PIMC) method allows one to
calculate, for a given interatomic potential, the equilibrium properties
of a bosonic system at finite temperature essentially without any
approximation.

The PIMC technique has been applied in the context of ultracold Bose gases
to investigate various thermodynamic  properties in harmonically confined
systems both in three~\cite{3dharm} and lower dimensions~\cite{ldharm}, and 
for a detailed study of the critical behavior and of the superfluid 
transition temperature in three-~\cite{3dhomo1,3dhomo2} and 
two-dimensional~\cite{2dhomo} homogeneous systems. 

In the present paper, we report on the results of a PIMC 
calculation of the
equation of state of a  three-dimensional homogeneous Bose gas as a
function of temperature and for different values of the interaction 
strength. The main focus is on the universal character exhibited by 
the equation of state in terms of the reduced temperature $T/T_c^0$, where
$T_c^0=(2\pi\hbar^2/m k_B)[n/\zeta(3/2)]^{2/3}$ is the BEC transition 
temperature of an ideal gas of particles of mass $m$ and number density
$n$, and of the gas parameter $na^3$,  incorporating the effects of
interatomic interactions at low temperatures through the $s$-wave
scattering length  $a$. 
We consider different repulsive model potentials
(hard sphere, soft sphere and negative-power potential) and we  explicitly
show the universal behavior of the energy per particle and pressure, both
below and above the transition  temperature, if the gas parameter is
small enough. At low temperatures, we compare the calculated 
energy per 
particle with the results of a Diffusion Monte Carlo (DMC) study carried
out at $T=0$\cite{US}, and at high  temperatures with the virial expansion
of an interacting gas. We believe that the present microscopic calculation 
could serve as a reference study for investigations of the thermodynamic 
properties of interacting Bose gases. 

The structure of the paper is as follows. In Sec.~\ref{Section2} we give a
brief overview of the PIMC method of  which we use two different
implementations: a fourth-order extension 
of the Trotter primitive
approximation  (for the negative-power potential) and the pair-product
approximation (for the hard- and soft-sphere potential).  In
Sec.~\ref{Section3} we discuss the results for the energy per particle and
gas pressure as a function of  temperature and interaction strength.
Finally in Sec.~\ref{Conclusions} we draw our conclusions.

\section{Method}
\label{Section2}

We consider a system of $N$ particles described by the following Hamiltonian
\begin{equation}
\hat{H}=- \frac{\hbar^2}{2m}\sum_{i=1}^N\nabla_i^2+\sum_{i<j}V(|{\bf r}_i-{\bf r}_j|) \;,
\label{hamiltonian}
\end{equation}
with different models for the spherical two-body interatomic 
potential $V(r)$: 

\noindent
1) Hard-sphere (HS) potential, defined by
\begin{equation}
V^{HS}(r)=\left\{  \begin{array}{cc} +\infty & (r<a)\;,  \\
                                   0    & (r>a)\;,  \end{array} \right.
\label{HS}
\end{equation}
for which the hard-sphere diameter $a$ corresponds to the $s$-wave scattering 
length.

\noindent
2) Soft-sphere (SS) potential, defined by    
\begin{equation}
V^{SS}(r)=\left\{  \begin{array}{cc}    V_0  & (r<R_0)\;,  \\
                                   0    & (r>R_0)\;,  \end{array} \right.
\label{SS}
\end{equation}
with $V_0>0$. In this case the scattering length is given by 
\begin{equation}
a=R_0 \left[1-\frac {\tanh(K_0 R_0)} {K_0 R_0} \right] \; , 
\label{aSS}
\end{equation}
with $K_0^2=V_0m/\hbar^2$. 
For finite $V_0$ one has always $R_0>a$, while for $V_0\to+\infty$ the SS
potential coincides  with the HS one with $R_0=a$. In the present
calculation the range of the SS potential is kept fixed to the value 
$R_0=5a$ and the height $V_0$ is determined to give the desired value of
$a$.

\noindent
3) Negative-power (NP) potential, defined by
\begin{equation}
V^{NP} (r)= \alpha/r^{p} \;,
\label{NP}
\end{equation}
with $\alpha>0$ and the integer $p>3$. 
For this potential the scattering length is given by~\cite{Landau}
\begin{equation}
a=\left( \frac{2m\alpha/\hbar^2}{(p-2)^2} \right)^{1/(p-2)}
\frac{\Gamma[(p-3)/(p-2)]}{\Gamma[(p-1)/(p-2)]} \;,
\label{aNP}
\end{equation} 
where $\Gamma(x)$ is the Gamma function. In the present calculation
we use $p=9$, which yields 
$a=(2m\alpha/49\hbar^2)^{1/7}\Gamma(6/7)/\Gamma(8/7)$. 

The universal regime in the plane $na^3$-$T/T_c^0$ is analyzed by
performing PIMC simulations using the above three potentials with the same
value for the gas parameter $na^3$.

The partition function $Z$ of a bosonic system with inverse temperature
$\beta=(k_BT)^{-1}$ is defined as the trace over all states of the density
matrix $\hat{\rho}=e^{-\beta \hat{H}}$ properly symmetrized. 
The partition function
satisfies the  convolution equation
\begin{eqnarray}
Z &=& \frac{1}{N!}\sum_P \int d{\bf R} \rho({\bf R},P{\bf R},\beta) = 
\frac{1}{N!}\sum_P \int d{\bf R} 
\label{PIMC1}\\ \nonumber
  &\times&  \int d{\bf R}_2 ... \int d{\bf R}_M \rho({\bf R},{\bf R}_2,\tau)...
  \rho({\bf R}_M,P{\bf R},\tau) \;,
\end{eqnarray}
where $\tau=\beta/M$, ${\bf R}$ collectively denotes the position vectors 
${\bf R}=({\bf r}_1,{\bf r}_2,...,{\bf r}_N)$, $P{\bf R}$ denotes the
position vectors with permuted labels  $P{\bf R}=({\bf r}_{P(1)},{\bf
r}_{P(2)},...,{\bf r}_{P(N)})$ and the sum extends over the $N!$
permutations of $N$  particles. In a PIMC calculation, one 
makes use of
suitable approximations for the density matrix  $\rho({\bf R},{\bf
R}^\prime,\tau)$ at the higher temperature $1/\tau$ in Eq. (\ref{PIMC1})
and performs the  multidimensional integration over ${\bf R}$, ${\bf
R}_2$,...,${\bf R}_M$ as well as the sum over permutations $P$ by  Monte
Carlo sampling~\cite{Ceperley}. The statistical expectation value of a
given operator $O({\bf R})$, 
\begin{equation}
\langle O\rangle = \frac{1}{Z}\frac{1}{N!}\sum_P 
\int d{\bf R} O({\bf R}) \rho({\bf R},P{\bf R},\beta) \;,
\label{PIMC2}
\end{equation}     
is calculated by generating stochastically a set of configurations $\{{\bf
R}_i\}$, sampled from a probability density  proportional to the
symmetrized density matrix, and then by averaging over the set of values
$\{O({\bf R}_i)\}$. 

Various approximations have been used for the density matrix at the high
effective temperature $M/\beta$. In a first approach, one relies on the
exact operator formula
\begin{equation}
e^{- \tau (\hat{T} + \hat{V}) +  \frac{\tau^2}{2} [\hat{T},\hat{V}]} =
e^{-\tau \hat{T}} e^{-\tau \hat{V}} \; ,
\label{rhoexact}
\end{equation} 
and approximates it in the limit $\tau \to 0$. The lowest order 
is known as primitive approximation (PA)
\begin{equation} 
e^{-\tau(\hat{T}+\hat{V})}= e^{-\tau \hat{V}/2}e^{-\tau \hat{T}}e^{-\tau
\hat{V}/2} + O(\tau^3) \;,
\label{PIMC3}
\end{equation}
and generate results for the energy with a quadratic dependence on $\tau$.
The convergence to the exact result is guaranteed by the Trotter
formula~\cite{Trotter},
\begin{equation}
e^{-\tau(\hat{T}+\hat{V})}= \lim_{M \to \infty} \left( e^{-\tau \hat{T}} 
e^{-\tau \hat{V}} \right)^M  \; ,
\label{trotter}
\end{equation}
but, from a practical point of view, PA is not accurate enough for studying
systems at temperatures below the superfluid transition~\cite{Brualla}. In order to
improve the accuracy of this approach, we have calculated the properties 
of the gas with the NP potential by means of a higher-order scheme based on
the operatorial decompositions proposed by Chin~\cite{chin}. Chin's
action (CA) is accurate to fourth-order in $\tau$ but
allows for a practical sixth-order dependence by adjusting some free
parameters of the decomposition~\cite{sakkos_future}.

An alternative approximation for the high temperature density matrix is
based on the pair-product ansatz (PPA)~\cite{Ceperley}
\begin{equation}
\rho({\bf R},{\bf R}^\prime,\tau)=\prod_{i=1}^N\rho_1({\bf r}_i,{\bf r}_
i^\prime,\tau)\prod_{i<j}
\frac{\rho_{rel}({\bf r}_{ij},{\bf r}_{ij}^\prime,\tau)}
{\rho_{rel}^0({\bf r}_{ij},{\bf r}_{ij}^\prime,\tau)} \;.
\label{PIMC4}
\end{equation}  
In the above equation $\rho_1$ is the single-particle ideal-gas density matrix
\begin{equation}
\rho_1({\bf r}_i,{\bf r}_i^\prime,\tau)=\left(\frac{m}{2\pi\hbar^2\tau}
\right)^{3/2} 
e^{-({\bf r}_i-{\bf r}_i^\prime)^2m/(2\hbar^2\tau)} \;,
\label{PIMC5}
\end{equation}
and $\rho_{rel}$ is the two-body density matrix of the
interacting system, which depends on 
the relative coordinates  ${\bf r}_{ij}={\bf r}_i-{\bf r}_j$ and ${\bf
r}_{ij}^\prime={\bf r}_i^\prime-{\bf r}_j^\prime$,  
divided by the corresponding ideal-gas term
\begin{equation}
\rho_{rel}^0({\bf r}_{ij},{\bf r}_{ij}^\prime,\tau)=
\left(\frac{m}{4\pi\hbar^2\tau}\right)^{3/2} 
e^{-({\bf r}_{ij}-{\bf r}_{ij}^\prime)^2 m/(4\hbar^2\tau)} \;.
\label{PIMC6}
\end{equation}
It can be shown~\cite{Ceperley} that PPA, Eq. (\ref{PIMC4}), is more
accurate than the simple PA, Eq. (\ref{PIMC3}), especially when the 
temperature of the system
is very low and the interactions are of hard-core type.
We have used the PPA approach for the simulations with the HS and SS
potentials which, in fact, can not be strictly used in the first approach
due to their discontinuous character.

The two-body density matrix at the inverse temperature $\tau$,
$\rho_{rel}({\bf r},{\bf r}^\prime,\tau)$, can be calculated  for a given
spherical potential $V(r)$ using the partial-wave decomposition 
\begin{eqnarray}
\rho_{rel}({\bf r},{\bf r}^\prime,\tau)&=&\frac{1}{4\pi}
\sum_{\ell=0}^\infty (2\ell +1)P_\ell(\cos\theta) 
\label{PIMC7} \\ \nonumber
&\times&\int_0^\infty dk e^{-\tau\hbar^2k^2/m} R_{k,\ell}(r)R_{k,\ell}(r^\prime) \;,
\end{eqnarray} 
where $P_\ell(x)$ is the Legendre polynomial of order $\ell$ 
and $\theta$ is the angle between ${\bf r}$ and ${\bf r}^\prime$.
The functions $R_{k,\ell}(r)$ are solutions of the radial 
Schr\"odinger equation 
\begin{eqnarray}
&-&\frac{\hbar^2}{m}\left( \frac{d^2R_{k,\ell}}{dr^2} +\frac{2}{r} \frac{dR_{k,\ell}}{dr} 
-\frac{\ell(\ell+1)}{r^2}R_{k,\ell}\right) \nonumber\\
&+& V(r)R_{k,\ell} = \frac{\hbar^2k^2}{m}R_{k,\ell} \;,
\label{PIMC8}
\end{eqnarray}
with the asymptotic behavior
\begin{equation}
R_{k,\ell}(r)=\sqrt{\frac{2}{\pi}}\frac{\sin(kr-\ell\pi/2+\delta_\ell)}{r} \;,
\label{PIMC9}
\end{equation}
holding for distances $r\gg R_0$, where $R_0$ is the range of the
potential. The phase shift $\delta_\ell$ of the  partial wave of order
$\ell$ is determined from the solution of Eq. (\ref{PIMC8}) for the given
interatomic potential $V(r)$. 

For the HS potential a simple analytical approximation of the
high-temperature two-body density matrix due to Cao and  Berne~\cite{Cao}
has been proven to be highly accurate. The Cao-Berne density matrix 
$\rho_{rel}^{CB}({\bf r},{\bf r}^\prime,\tau)$ is obtained using the large
momentum expansion of the HS phase shift  $\delta_\ell\simeq-ka+\ell\pi/2$
and the large momentum expansion of the solutions of the Sch\"odinger
equation (\ref{PIMC8}) $R_{k,\ell}(r)\simeq\sqrt{2/\pi}\sin[k(r-a)]/r$.
This yields the result
\begin{eqnarray}
\frac{\rho_{rel}^{CB}({\bf r},{\bf r}^\prime,\tau)}
{\rho_{rel}^0({\bf r},{\bf r}^\prime,\tau)}&=& 
 1 -\frac{a(r+r^\prime)-a^2}{rr^\prime} \\ \nonumber
&\times& e^{-[rr^\prime +a^2-a(r+r^\prime)](1+\cos\theta)m/(2\hbar^2\tau)} \;.
\label{CaoBerne}
\end{eqnarray}

In the case of the SS potential, we calculate numerically 
$\rho_{rel}({\bf
r},{\bf r}^\prime,\tau)$ from  Eqs. (\ref{PIMC7})-(\ref{PIMC9}). In the
case of the HS potential, we use both the density matrix determined 
numerically and the Cao-Berne approximation [Eq. (\ref{CaoBerne})]. We
have verified that in all cases the two  procedures give indistinguishable
results for the HS interaction within the present statistical uncertainty. 

\section{Results}
\label{Section3}
 
PIMC simulations have been carried out for a Bose gas with periodic
boundary conditions and with $N$ ranging from  64 to 1024. In all the
calculations finite size effects have been checked to be smaller than the
reported statistical uncertainty.

\subsection{Normal phase}

\begin{figure}
\begin{center}
\includegraphics*[width=8.5cm]{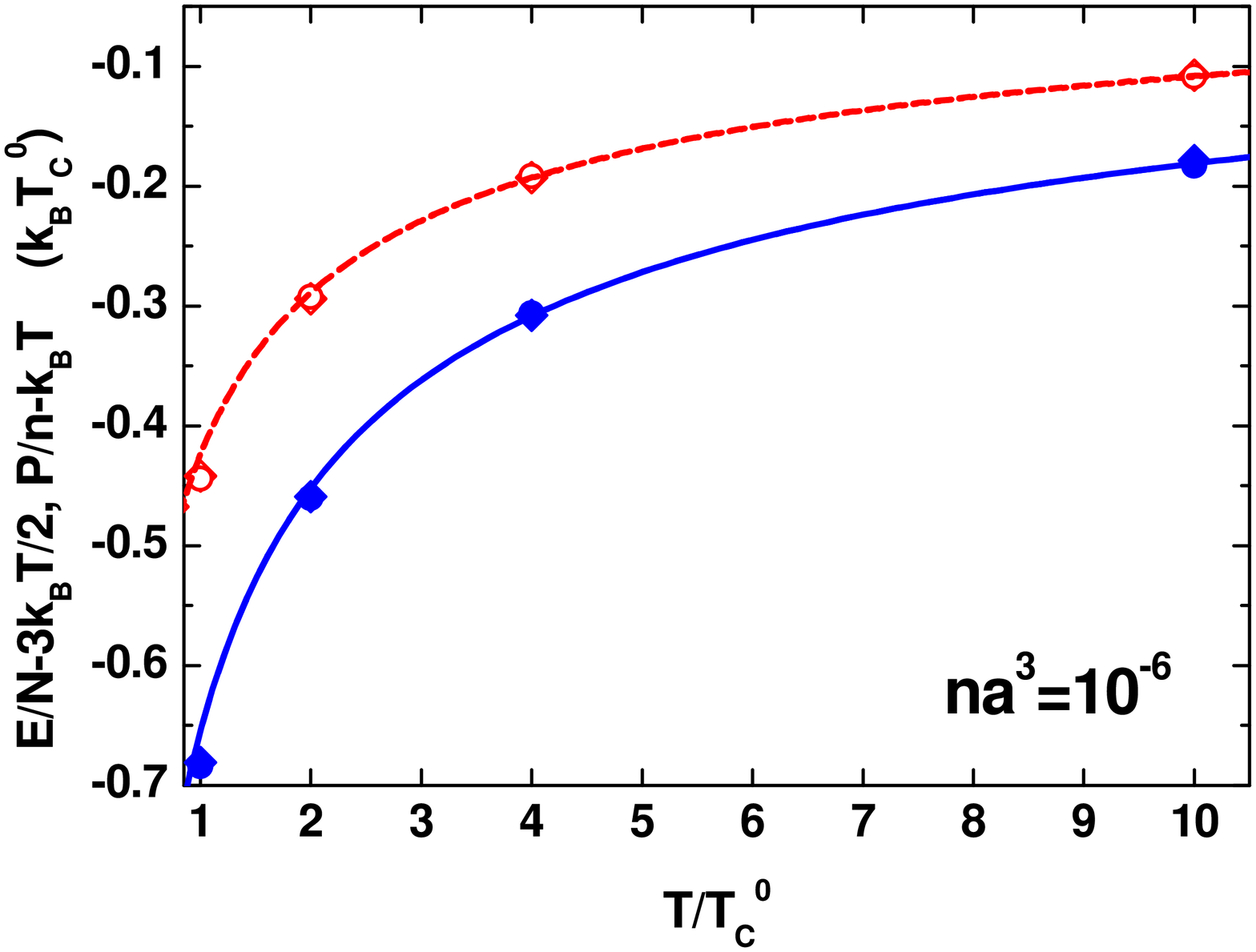}
\caption{(color online). Energy per particle and pressure of a Bose gas in
the normal phase as a function of  temperature. The gas parameter is
$na^3=10^{-6}$. Solid symbols (blue online) refer to $E/N-3k_BT/2$: HS
potential  (circles), SS potential (diamonds). Open symbols (red online)
refer to $P/n-k_BT$: HS potential (circles), SS potential  (diamonds).
Statistical error bars are smaller than the size of the symbols. The virial
expansion (\ref{virial2}) is  represented by lines (blue online): HS
potential (solid line), SS potential (long-dashed line). The virial
expansion  (\ref{virial1}) is represented by lines (red online): HS
potential (short-dashed line), SS potential (dotted line).}
\label{fig1}
\end{center}
\end{figure}

\begin{figure}
\begin{center}
\includegraphics*[width=8.5cm]{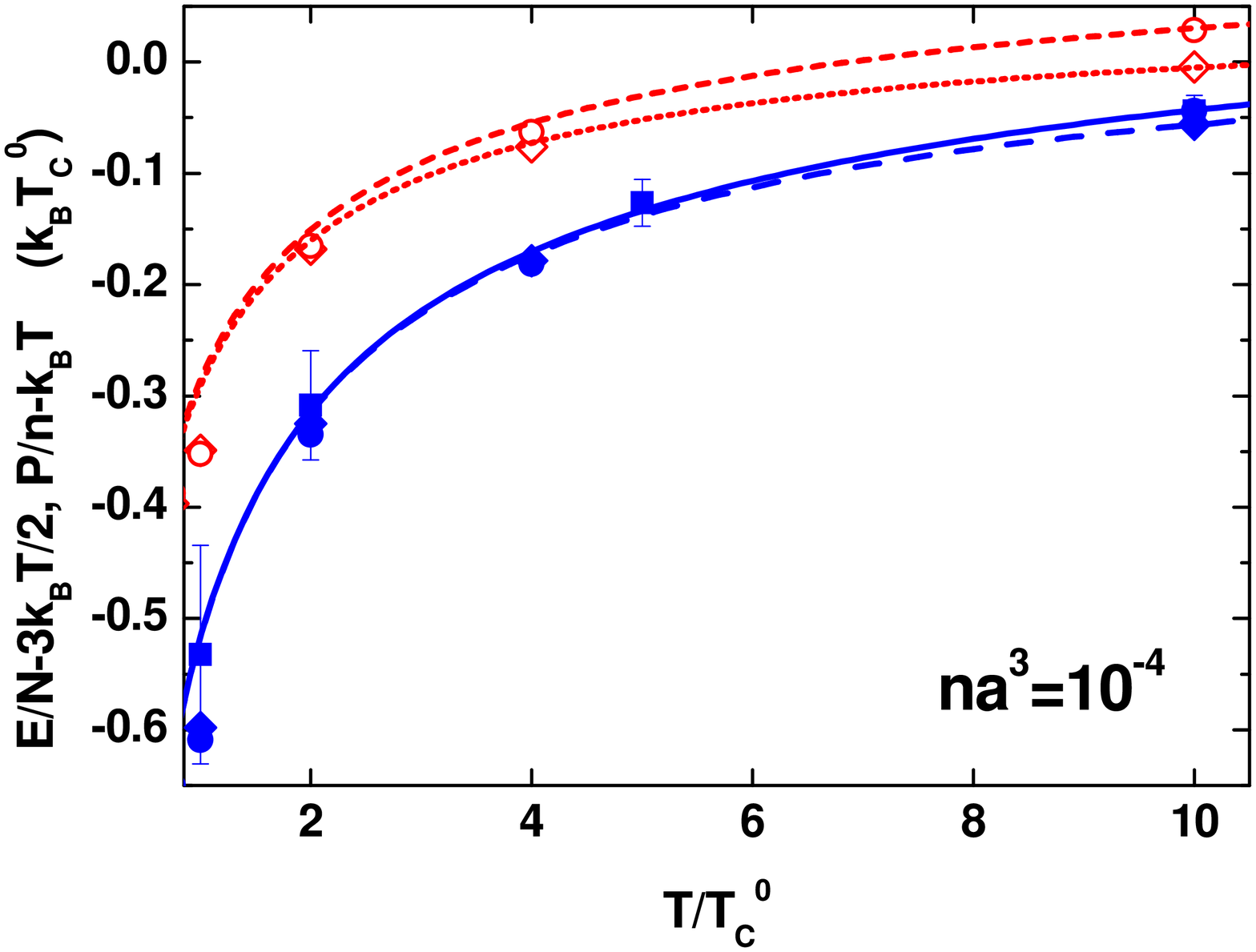}
\caption{(color online). Energy per particle and pressure of a Bose gas in
the normal phase as a function of  the temperature for $na^3=10^{-4}$. Same
notation as in Fig.~\ref{fig1}, except for the results of the energy 
for the NP potential, shown here as squares (blue online).}    
\label{fig2}
\end{center}
\end{figure}

\begin{figure}
\begin{center}
\includegraphics*[width=8.5cm]{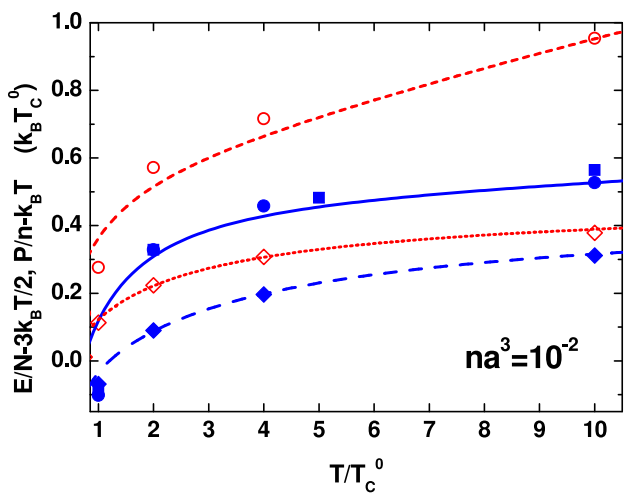}
\caption{(color online). Energy per particle and pressure of a Bose gas in
the normal phase as a function of  the temperature for $na^3=10^{-2}$. Same
notation as in Figs.~\ref{fig1} and \ref{fig2}.}
\label{fig3}
\end{center}
\end{figure}

At high temperatures $n\lambda_T^3\ll 1$, where
$\lambda_T=\sqrt{2\pi\hbar^2/mk_BT}$ is the thermal wave length, the 
equation of state of the gas can be calculated from the virial expansion
\begin{equation}
\frac{PV}{Nk_BT}=1+a_2(T)n\lambda_T^3+... \;, 
\label{virial1}
\end{equation}
where we considered only the contribution arising from the second virial
coefficient $a_2(T)$. The corresponding virial  expansion of the energy per
particle can be calculated using standard thermodynamic relations and one
finds~\cite{Landau1} 
\begin{equation}
\frac{E}{N}=\frac{3}{2}k_BT \left( 1+a_2(T)n\lambda_T^3+...\right) \;. 
\label{virial2}
\end{equation}
For a non-interacting Bose gas the second virial coefficient can be
promptly calculated with the result  $a_2^0=-1/\sqrt{2^5}$, determined by
statistical effects. For a gas of particles interacting through a
repulsive  interatomic potential $a_2(T)$ can be calculated through a
summation over partial waves~\cite{Pathria}
\begin{eqnarray}
a_2(T) = a_2^0 &-& \frac{\sqrt{8}}{\pi} \sum_{\ell=0,2,4,...} (2\ell+1)
\nonumber\\
&\times& \int_0^\infty dk e^{-\hbar^2k^2/mk_BT} 
\frac{\partial\delta_\ell(k)}{\partial k} \;.
\label{virial3}
\end{eqnarray}
For bosons, the sum in Eq.~(\ref{virial3}) only includes 
even partial waves. The $\ell$-th
partial-wave phase shift $\delta_\ell(k)$ in the  above equation is
obtained  from the solution of the
Schr\"odinger equation (\ref{PIMC8}) for the given potential $V(r)$
with  the boundary condition
(\ref{PIMC9}). If the thermal wave length is much larger than the range of
the potential, $\lambda_T\gg R_0$, one obtains, to lowest order,  
the following result
\begin{equation}
a_2(T)-a_2^0=2\frac{a}{\lambda_T}+... \;,
\label{virial4}
\end{equation} 
which only depends on the scattering length $a$.

In Figs.~\ref{fig1}-\ref{fig3}, we present the PIMC results 
for the energy per
particle $E/N$ and pressure $P$ in  the normal phase ($T>T_c$) for three
different values of the gas parameter: $na^3=10^{-6}$ (Fig.~\ref{fig1}), 
$na^3=10^{-4}$ (Fig.~\ref{fig2}) and $na^3=10^{-2}$ (Fig.~\ref{fig3}). In
order to emphasize the deviations from the  classical results we plot the
quantities $E/N-3k_BT/2$ and $P/n-k_BT$. We also  plot the corresponding
virial expansions from Eqs.~(\ref{virial1}),(\ref{virial2}) with the second
virial coefficient  $a_2(T)$ calculated using Eq.~(\ref{virial3}) for the
HS and SS potentials. For the smallest value of the interaction  
strength,
$na^3=10^{-6}$, we find very good agreement between the HS and SS results
and with the virial expansions  down to temperatures close to the
transition temperature. At $na^3=10^{-4}$, the virial expansion still
provides a good  approximation in the whole temperature regime and
deviations are found only at the lowest temperatures $T\sim T_c^0$. 
On the other hand, universality is maintained
only for low $T$ since differences between the HS
and SS potentials start to become visible for temperatures 
$T/T_c^0\gtrsim 4$. Finally, for the largest interaction strength
$na^3=10^{-2}$,  the universal behavior fixed by the scattering length $a$
is lost in the whole temperature range. For the HS potential,
agreement
with the virial expansion is found only at the largest temperature.
Concerning the results for the  NP potential obtained with the CA
approximation,
notice that already for $na^3=10^{-4}$ the statistical uncertainty is 
significantly larger than the one corresponding to results for the HS and
SS potentials obtained using the PPA. This fact is due to the large
separation in scale between the range of interactions and the mean
interparticle distance which  occurs in dilute systems. For very small
values of the gas parameter $na^3$ the algorithm based on the PPA, which
is  constructed from the exact solution of the two-body problem, converges
much faster than the one based on the CA.  For example, at
$T=2T_c^0$ and $na^3=10^{-4}$, the calculation for the NP potential has
been performed using up to  $M=200$ beads in contrast to only
$M=12$ in the case of the HS and SS potentials. For the smallest value of
the  interaction  strength, $na^3=10^{-6}$, and especially for temperatures
below the transition temperature, the calculation  using the CA  
approach becomes much more computationally demanding due to the large
number of beads required.

\subsection{Superfluid phase}

\begin{figure}
\begin{center}
\includegraphics*[width=8.5cm]{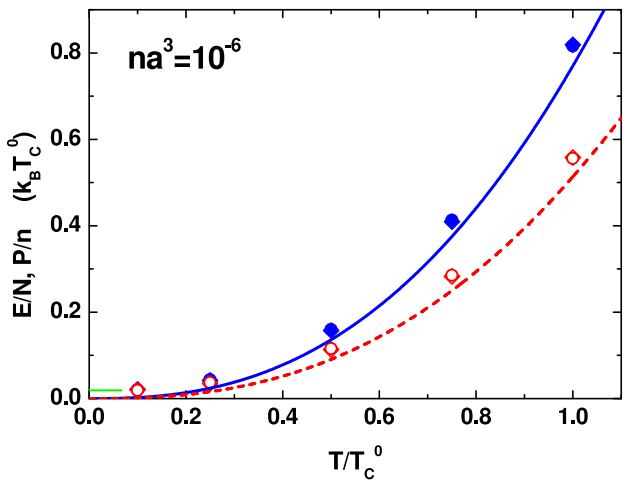}
\caption{(color online). Energy per particle and pressure of a Bose gas in
the superfluid phase as a function  of temperature. The gas parameter is
$na^3=10^{-6}$. Solid symbols (blue online) refer to $E/N$: 
HS potential 
(circles), SS potential (diamonds). Open symbols (red online) refer to
$P/n$: HS potential (circles), SS potential  (diamonds). Statistical error
bars are smaller than the size of the symbols. The horizontal bar (green
online)  corresponds to the ground-state energy per particle $E_0/N$ of a
HS gas calculated using DMC. The lines correspond  to a non-interacting
gas: the solid line (blue online) refers to the energy 
per particle and the
dashed line  (red online) to the pressure.}
\label{fig4}
\end{center}
\end{figure}

\begin{figure}
\begin{center}
\includegraphics*[width=8.5cm]{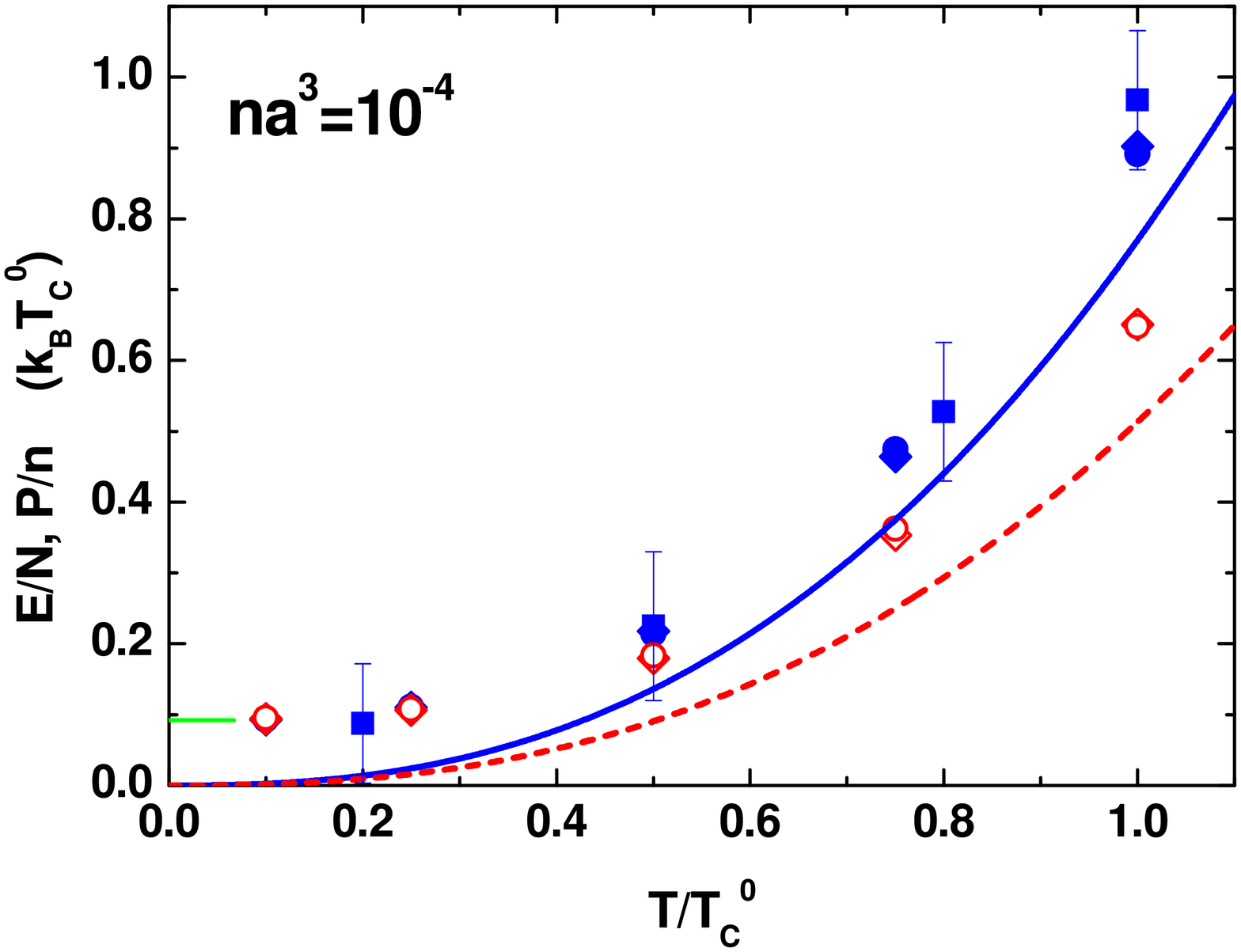}
\caption{(color online). Energy per particle and pressure of a Bose gas in
the superfluid phase as a function  of the temperature
for $na^3=10^{-4}$. Same
notation as in Fig.~\ref{fig4}, except for the results of the energy 
for the NP potential, shown here as squares (blue online).}
\label{fig5}
\end{center}
\end{figure}

\begin{figure}
\begin{center}
\includegraphics*[width=8.5cm]{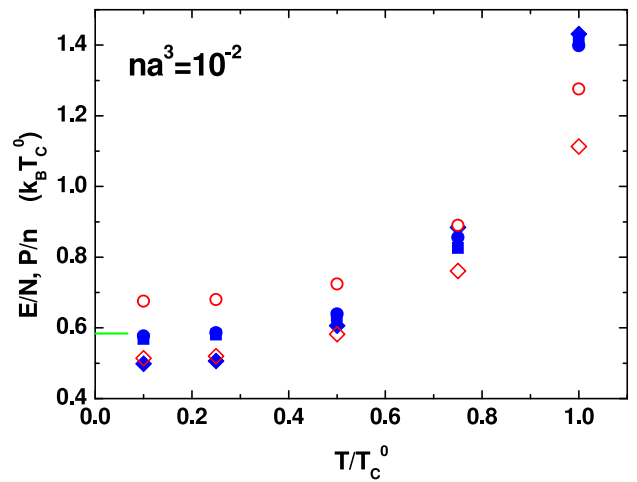}
\caption{(color online). Energy per particle and pressure of a Bose gas in
the superfluid phase as a function of  the temperature for $na^3=10^{-2}$.
 Same notation as in Figs.~\ref{fig4} and \ref{fig5}.}
\label{fig6}
\end{center}
\end{figure}

The determination of the transition temperature from the equation of state
is very delicate as it involves a slight change  of the energy vs. $T$
dependence at $T_c$. Its precise determination would require the
calculation of the specific heat. Other observables, 
like the superfluid density and the condensate  fraction, give a direct 
signature of the
transition~\cite{3dhomo1,3dhomo2}. The most reliable results so far give
the  following shift of the transition temperature $T_c$~\cite{3dhomo2}
\begin{equation} 
\frac{T_c-T_c^0}{T_c^0}=(1.29\pm 0.05) (na^3)^{1/3} \;,
\label{deltaTc} 
\end{equation}    
holding for very small values of the gas
parameter. For the values of $na^3$ used in the present study, the above
equation  yields estimates of $T_c$ ranging from 1\% to 30\% above $T_c^0$.
We are not interested here in the calculation of $T_c$  and the focus is
only on the precise determination of the equation of state.

In Figs.~\ref{fig4}-\ref{fig6}, we show results for 
$E/N$ and $P/n$ in the
superfluid phase ($T<T_c$) for the three values  of the gas parameter:
$na^3=10^{-6}$ (Fig.~\ref{fig4}), $na^3=10^{-4}$ (Fig.~\ref{fig5}) and
$na^3=10^{-2}$ (Fig.~\ref{fig6}).  To emphasize the effects of interactions
in Figs.~\ref{fig4} and \ref{fig5} we also plot the energy per particle and
the  pressure of an ideal Bose gas. For $na^3=10^{-6}$ and $na^3=10^{-4}$
we see very small differences between the results of the HS and SS
potentials (notice the large statistical uncertainty in the results of the
NP potential at $na^3=10^{-4}$). For the  largest value of $na^3$ we still
find good agreement between the results of the HS and NP potential, while
significant  differences are found between the HS and SS potential. At very
low temperatures, the PIMC results agree with the 
ground-state  energy per
particle $E_0/N$ obtained for the HS potential using the DMC
method~\cite{US}. For the three values of the gas  parameter used in the
present study the results are (in units of $k_BT_c^0$): $E_0/N=0.01905(2)$
at $na^3=10^{-6}$,  $E_0/N=0.09185(8)$ at $na^3=10^{-4}$, and
$E_0/N=0.5840(4)$ at $na^3=10^{-2}$.     

The PIMC results for $E/N$ and $P/n$ obtained using the HS and SS potential
at the various temperatures and for the three  values of $na^3$ are
reported in Tables~\ref{tab1}, \ref{tab2}. Finite size effects are relevant
only in the simulations  performed at $T=T_c^0$ because of the vicinity to
the critical point. For example, in the case of the HS potential at 
$na^3=10^{-4}$ we find for this temperature the result:
$E/(Nk_BT_c^0)=0.935(7)$ with $N=64$ and $E/(Nk_BT_c^0)=0.891(13)$  with
$N=1024$.

\section{Conclusions}
\label{Conclusions}

In conclusion, we have carried out using exact PIMC methods a precision 
calculation of the equation of state of an interacting Bose gas as a 
function of temperature and interaction strength. The universal character 
of the equation of state at low temperatures and small values of the gas 
parameter is pointed out by performing simulations with different interatomic 
model potentials. Above the transition temperature we compare our results for 
energy and pressure with the high-temperature expansion based on the second virial
coefficient. The inclusion of tables for both energy and pressure is intended 
as a reference for future studies of the thermodynamic properties of interacting 
Bose gases.

\acknowledgments
SP and SG acknowledge support by the Ministero dell'Istruzione,
dell'Universit\`a e della Ricerca (MIUR). KS, JB and  JC acknowledge
support from DGI (Spain) Grant No. FIS2005-04181 and Generalitat de
Catalunya Grant No. 2005SGR-00779.  


\begin{table*}
\centering
\begin{ruledtabular}
\caption{Energy per particle $E/N$ (in units of $k_BT_c^0$) for the HS and 
SS potential and different values of the
gas parameter $na^3=10^{-6}$, 10$^{-4}$, 10$^{-2}$. In parenthesis we give 
the statistical errors.}
\begin{tabular}{ccccccc}
 & \multicolumn{6}{c}{$E/N$  $\;\;\;\;\;\;(k_BT_c^0$)} \\
\cline{2-7}
 & \multicolumn{2}{c}{$na^3=10^{-6}$} & \multicolumn{2}{c}{$na^3=10^{-4}$} & \multicolumn{2}{c}{$na^3=10^{-2}$}\\  
\cline{2-7}
$T/T_c^0$  & $HS$ & $SS$ & $HS$ & $SS$ & $HS$ & $SS$ \\
\hline
0.10  &  0.022(1)   &  0.021(1)    &  0.093(1)   &  0.093(1)    &  0.577(5)    &  0.499(1)  \\
0.25  &  0.043(2)   &  0.042(1)    &  0.111(2)   &  0.110(1)    &  0.586(4)    &  0.506(1)  \\
0.50  &  0.160(4)   &  0.158(3)    &  0.214(4)   &  0.218(4)    &  0.639(5)    &  0.606(4)  \\
0.75  &  0.402(5)   &  0.409(7)    &  0.475(14)  &  0.464(13)   &  0.856(13)   &  0.885(13) \\
1.00  &  0.816(10)  &  0.819(6)    &  0.891(13)  &  0.902(15)   &  1.398(10)   &  1.431(18) \\
2.00  &  2.540(6)   &  2.541(5)    &  2.665(13)  &  2.675(8)    &  3.328(4)    &  3.090(3)  \\
4.00  &  5.694(5)   &  5.692(5)    &  5.818(7)   &  5.821(5)    &  6.458(3)    &  6.196(3)  \\
10.0  &  14.817(5)  &  14.821(5)   &  14.956(4)  &  14.943(3)   &  15.527(6)   &  15.312(4) \\
20.0  &  29.885(5)  &  29.880(5)   &  30.023(4)  &  29.989(2)   &  30.615(11)  &  30.374(4) \\
\label{tab1}
\end{tabular}
\end{ruledtabular}
\end{table*} 

\begin{table*}
\centering
\begin{ruledtabular}
\caption{Pressure $P/n$ (in units of $k_BT_c^0$) for the HS and SS potential and different values of the
gas parameter $na^3=10^{-6}$, 10$^{-4}$, 10$^{-2}$. In parenthesis we give the statistical errors.}
\begin{tabular}{ccccccc}
 & \multicolumn{6}{c}{$P/n$  $\;\;\;\;\;\;(k_BT_c^0$)} \\
\cline{2-7}
 & \multicolumn{2}{c}{$na^3=10^{-6}$} & \multicolumn{2}{c}{$na^3=10^{-4}$} & \multicolumn{2}{c}{$na^3=10^{-2}$}\\  
\cline{2-7}
$T/T_c^0$  & $HS$ & $SS$ & $HS$ & $SS$ & $HS$ & $SS$ \\
\hline
0.10  &  0.021(1)   &  0.021(1)   &  0.095(1)   &  0.094(1)   &  0.676(6)    &  0.514(1)  \\
0.25  &  0.037(2)   &  0.035(1)   &  0.107(2)   &  0.106(1)   &  0.680(4)    &  0.520(2)  \\
0.50  &  0.116(3)   &  0.114(2)   &  0.183(4)   &  0.180(4)   &  0.724(4)    &  0.582(3)  \\
0.75  &  0.285(5)   &  0.283(5)   &  0.362(9)   &  0.353(9)   &  0.890(8)    &  0.761(8)  \\
1.00  &  0.556(7)   &  0.558(4)   &  0.648(10)  &  0.651(10)  &  1.276(8)    &  1.113(12) \\
2.00  &  1.708(4)   &  1.706(3)   &  1.835(9)   &  1.832(5)   &  2.572(3)    &  2.223(3)  \\
4.00  &  3.808(4)   &  3.807(3)   &  3.937(5)   &  3.924(4)   &  4.716(4)    &  4.308(3)  \\
10.0  &  9.891(4)   &  9.893(3)   &  10.028(3)  &  9.995(2)   &  10.954(8)   &  10.378(3) \\
20.0  &  19.936(3)  &  19.932(3)  &  20.075(4)  &  20.024(2)  &  21.335(10)  &  20.416(3) \\
\label{tab2}
\end{tabular}
\end{ruledtabular}
\end{table*}


\begin{thebibliography}{20}

\bibitem{BEC} M.H. Anderson {\it et al.}, Science {\bf 269}, 198 (1995); K.B. Davis {\it et al.}, Phys. Rev. Lett.
{\bf 75}, 3969 (1995); C.C. Bradley {\it et al.}, Phys. Rev. Lett. {\bf 75}, 1687 (1995). 

\bibitem{book} L.P. Pitaevskii and S. Stringari, {\it Bose-Einstein Condensation}, (Clarendon Press, Oxford, 2003).

\bibitem{review} F. Dalfovo, S. Giorgini, L.P. Pitaevskii and S. Stringari, Rev. Mod. Phys. {\bf 71}, 463 (1999).

\bibitem{3dharm} W. Krauth, Phys. Rev. Lett. {\bf 77}, 3695 (1996); M. Holzmann, W. Krauth and M. Naraschewski, 
Phys. Rev. A {\bf 59}, 2956 (1999); M. Holzmann and Y. Castin, Eur. Phys. J. D {\bf 7}, 425 (1999).

\bibitem{ldharm} S. Heinrichs, W.J. Mullin, J. Low Temp. Phys. {\bf 113}, 231 (1998); K. Nho and D. Blume, Phys. 
Rev. Lett. {\bf 95}, 193601 (2005).

\bibitem{3dhomo1} P. Gr\"uter, D.M. Ceperley and F. Lalo\"e, Phys. Rev. Lett. {\bf 79}, 3549 (1997); M. Holzmann 
and W. Krauth, Phys. Rev. Lett. {\bf 83}, 2687 (1999).

\bibitem{3dhomo2} V. A. Kashurnikov, N.V. Prokof'ev and B.V. Svistunov, Phys. Rev. Lett. {\bf 87}, 120402 (2001); 
N. Prokof'ev, O. Ruebenacker and B. Svistunov, Phys. Rev. A {\bf 69}, 053625 (2004); K. Nho and D.P. Landau, 
Phys. Rev. A {\bf 70}, 053614 (2004).

\bibitem{2dhomo} N. Prokof'ev, O. Ruebenacker and B. Svistunov, Phys. Rev. Lett. {\bf 87}, 270402 (2001).

\bibitem{US} S. Giorgini, J. Boronat and J. Casulleras, Phys. Rev. A {\bf 60}, 5129 (1999).

\bibitem{Landau} L.D. Landau and E.M. Lifshitz, {\it Quantum Mechanics} (Non-relativistic Theory) (Pergamon Press,
Oxford, 1977), pag. 550.

\bibitem{Ceperley} E.L. Pollock and D.M. Ceperley, Phys. Rev. B {\bf 30}, 2555 (1984); {\it ibid.} {\bf 36}, 8343 
(1987); D.M. Ceperley, Rev. Mod. Phys. {\bf 67}, 1601 (1995).

\bibitem{Trotter}  H.F. Trotter, Proc. Am. Math. Soc. {\bf 10}, 545 (1959).

\bibitem{Brualla} L. Brualla, K. Sakkos, J. Boronat and J. Casulleras, J. Chem. Phys. {\bf 121}, 636 (2004).

\bibitem{chin} S. A. Chin and C. R. Chen, J. Chem. Phys. \textbf{117}, 1409
(2002).

\bibitem{sakkos_future} K. Sakkos, J. Casulleras, and J. Boronat, to be published.

\bibitem{Cao} J. Cao and B.J. Berne, J. Chem. Phys. {\bf 97}, 2382 (1992).

\bibitem{Landau1} L.D. Landau and E.M. Lifshitz, {\it Statistical Physics} Part 1 (3rd edition) (Pergamon Press,
Oxford, 1980), Ch. 7.    

\bibitem{Pathria} R.K. Pathria, {\it Statistical Mechanics} 2nd edition (Butterworth-Heinemann, Oxford, 1996), 
pag. 252-253.

\end{thebibliography}
\end{document}